\documentclass{elsart}

\usepackage{natbib}

\newcommand{\aap}{A\&A}

\newcommand{\aj}{AJ}
\newcommand{\apj}{ApJ}
\newcommand{\apjl}{ApJ}
\newcommand{\apjs}{ApJ}

\newcommand{\mnras}{MNRAS}
\newcommand{\nat}{Nature}
\newcommand{\procspie}{Proc. SPIE}

\usepackage{graphicx}
\usepackage{amssymb}

\begin{document}

\begin{frontmatter}

\title{Witnessing the formation of a brightest cluster galaxy at ${\bf z>2}$}
\author{Roderik A. Overzier}
\ead{overzier@pha.jhu.edu}
\address{The Johns Hopkins University, 3400 N. Charles Street, Baltimore, MD 21218}
\author{George K. Miley}
\address{University of Leiden, Leiden Observatory, Postbus 9513, 2300 RA Leiden, The Netherlands}
\author{Holland C. Ford}
\address{The Johns Hopkins University, 3400 N. Charles Street, Baltimore, MD 21218}

\begin{abstract}
We present deep observations taken with the HST Advanced Camera for Surveys of the central massive galaxy in a forming cluster at
$z=2.2$. The galaxy hosting the powerful radio source MRC 1138--262 is associated with one of the most extensive merger 
systems known in the early universe. Our HST/ACS image shows many star-forming galaxies merging within a $\sim$200 
kpc region that emits both diffuse line emission and continuum in the rest-frame UV. Because this galaxy lives in an 
overdense environment, it represents a rare view of a brightest cluster galaxy in formation at $z>2$ which may serve as a 
testbed for predictions of massive cluster galaxy formation.
\end{abstract}


\end{frontmatter}

\section{Introduction}

The formation and evolution of structure in the universe is a
fundamental field of research in  
cosmology. Clusters of galaxies represent the most extreme deviation from initial conditions, 
and are therefore good evolutionary probes for studying the  
formation of the large-scale structure. While clusters of galaxies have been studied extensively in the 
relatively nearby universe, their evolutionary history becomes obscure beyond roughly 
half the Hubble time (e.g. Blakeslee et al. 2006, Mullis et al. 2005, Stanford et al. 2006).   
Their progenitors are difficult to identify when the density contrast between the 
forming cluster and the field becomes small, and mass condensations on the scales of clusters 
are extremely rare at any epoch (Kaiser 1984). 
While deep pencil beam surveys have allowed us to study many different kinds of galaxies and their relative contributions 
to the cosmic star formation rate out to $z\sim7-8$, the study of the evolution of galaxy clusters has not progressed 
much beyond $z\sim1$. However, the relatively advanced evolutionary state of some clusters even at $z=1$ suggests a formation 
epoch at $z>2$, consistent with predictions for the (early) formation of large-scale structure in cosmological 
simulations (Springel et al. 2005). Finding and studying the progenitors of these clusters will give new clues to how 
the most massive structures in the Universe came about. 

Distant radio galaxies are important laboratories for studying the formation and evolution of massive galaxies in the early universe. 
They are among the most luminous and largest known galaxies at $z\gtrsim1$, having stellar masses in excess of 
$10^{11}$ $M_\odot$ (e.g. De Breuck et al. 2002, Villar-Martin et al. 2006). 
There is evidence that protoclusters in the early Universe ($2<z<5$) can be found around (but are not limited to) 
radio galaxies (e.g. Kurk 2000, 2004; 
Miley et al. 2004; Overzier et al. 2006a,2006b; Pentericci et al. 2000,2002; Venemans et al. 2002, 2004, 2005). 
These protoclusters have galaxy and mass overdensities that are sufficiently 
large for such structures to break away from the expanding Universe, and form a virialized cluster before $z=0$. 
Note that although powerful 3C-type radio 
sources are very rare objects, their number density is comparable to that of Abell clusters ($\sim10^{-6}$--$10^{-5}$  Mpc$^{-3}$) 
if estimated radio source life-times ($\lesssim10^8$ yr) are taken into account (Bahcall \& Soneira 1983, Dunlop \& Peacock 1990, West 1994).

It is believed that the fueling of a supermassive black hole due to cooling flows, mergers and interactions are responsible for triggering 
a powerful (radio) quasar (e.g. Heckman et al. 1986, Kauffmann \& Haehnelt 2000, Canalizo \& Stockton 2001, Croton et al. 2006). 
If this indeed provides the trigger for radio activity, then the cores of protoclusters are likely locations to find radio galaxies at high 
redshift. Taken together with the highly special intrinsic properties of the host galaxies of distant luminous radio sources, it is probable 
that radio galaxies could be progenitors of the brightest cluster galaxies (BCGs). 
The characteristic luminosity of radio sources increases with redshift in a flux limited sample, meaning that one could be observing an 
increasing fraction of proto-BCGs with increasing redshift, while at low redshift ($z<0.5$) 3C sources span a factor of $10^4$ in radio 
luminosity and most of these will therefore not correspond to BCGs (West et al. 1994). 

\section{HST/ACS observations of a proto-BCG at ${\bf z=2.2}$}

We obtained deep images of the radio galaxy MRC 1138--262 with the {\it Hubble Space Telescope} 
Advanced Camera for Surveys (ACS, Ford et al. 1998), as part of an ACS GTO programme to map 
the central regions of radio galaxy protoclusters in order to study star formation and 
galaxy morphologies in overdense environments at $2<z<6$. 
By coadding the images taken through the filters $g_{475}$ and $I_{814}$ (both in the rest-frame UV) 
we created a 44 ksec image of a $100\times100$ kpc$^2$ region surrounding the radio galaxy (Fig. 1). 
Initially studied in detail by Pentericci et al. (1997,1998,2001), the increased depth of the ACS 
image reveals more fine details and complexity in this remarkable object (Miley et al. 2006):

\begin{figure}[t]
\begin{center}
\includegraphics[width=\textwidth]{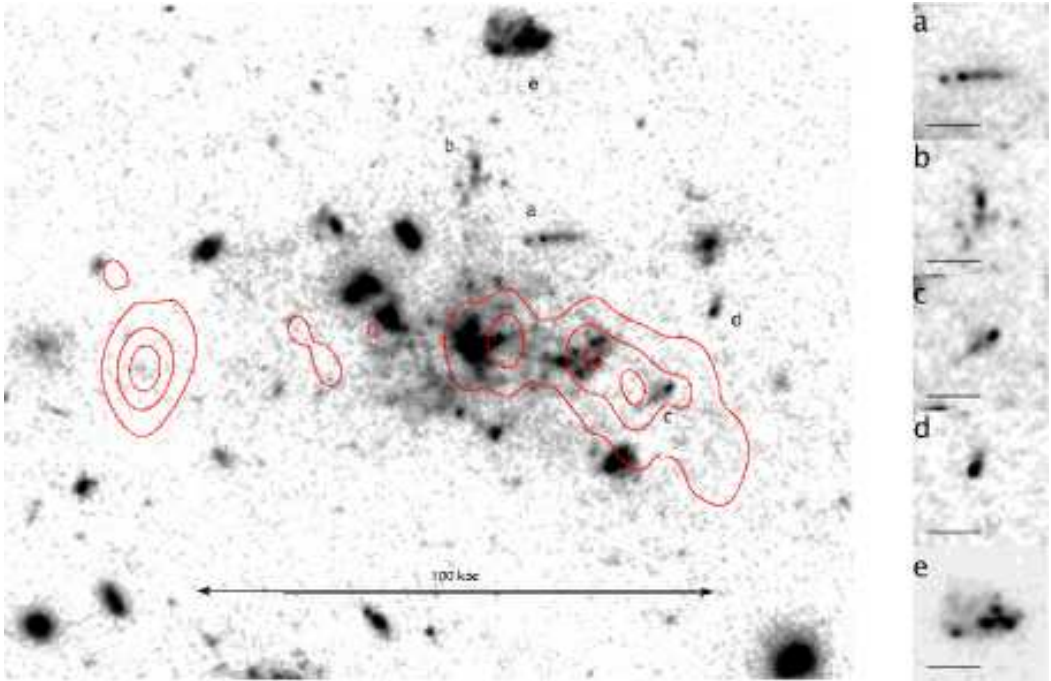}
\end{center}
\caption{HST/ACS $g_{475}+I_{814}$ image of the radio galaxy MRC 1138--262 at $z=2.2$. 
The system consists of a large ($>100$ kpc) conglomeration of sub-galactic clumps embedded in a region of diffuse emission. 
The total star formation rate in the clumps and the diffuse component (several hundred $M_\odot$ yr$^{-1}$) are about equal. 
Comparison with the rest-frame optical magnitude (Pentericci et al. 1997) indicates that the mass of the star-forming component 
seen in this image is only a small fraction of the total mass of the system, suggesting that the MRC 1138--262 system has 
elements of both early and late formation (see Miley et al. 2006 for details). Contours show the radio emission at 4.5 GHz. 
The postage stamps to the right of the image show close-ups of objects having interesting morphologies 
(marked with the letters a--e in the image).}
\end{figure}

The optical continuum emission of the galaxy consists of at least 10 distinct clumps, presumably satellite 
galaxies that are still merging with MRC 1138--262. Several of the faint satellites have interesting morphologies. 
Object `a' in Fig. 1 shows four bright knots of emission in a line like a string of beads. These longitudinal galaxies that are 
barely resolved in their transverse dimension are known as `chain' galaxies, and could be distant, 
ancestral forms of local Hubble type galaxies (Cowie et al. 1995). They are believed to form from star formation 
progressing along filaments of gas previously blown out from massive forming galaxies (Taniguchi \& Shioya 2001). 
Because the chain galaxies are believed to be inherently unstable and therefore short-lived, the chance of finding one 
could have been enhanced by the vigorous merging and gaseous outflows observed in the MRC 1138--262 system. The objects marked 
`c' and `d' in Fig. 1 can be classified as `tadpole' galaxies, which consist of a knot at one end and an extended tail at the other. These objects are believed to be early-stage mergers (Straughn et al. 2006). Other objects also show evidence for disruption, 
multiple nuclei, and tidal tails. 

Interestingly, faint diffuse emission is visible throughout the $\sim100$ kpc inter-clump region. 
The extended emission is unlikely to be caused by scattered light of the obscured quasar nucleus, because its 
roughly spherical morphology and radial profile do not resemble that of a scattering cone. The mean colour of the diffuse emission 
is comparable to that of the star-forming satellites, consistent with the occurrence of ongoing star formation over a very extended region. 
The small-scale variations in the surface brightness of the extended emission seen to the south-west of the nucleus could be 
caused by gradients in the amount of dust, or by the subclustering of star-forming clumps too faint to be detected individually. 
The total extended luminosity (comprising 45\% of the total emission in $g_{475}$) corresponds to a star formation rate 
of $\gtrsim$100 $M_\odot$ yr$^{-1}$. This indicates that star formation over extended regions such as observed in MRC 1138--262 
could be important for building the most massive galaxies, besides merging and gas accretion.

\begin{figure}[t]
\begin{center}
\includegraphics[width=\textwidth]{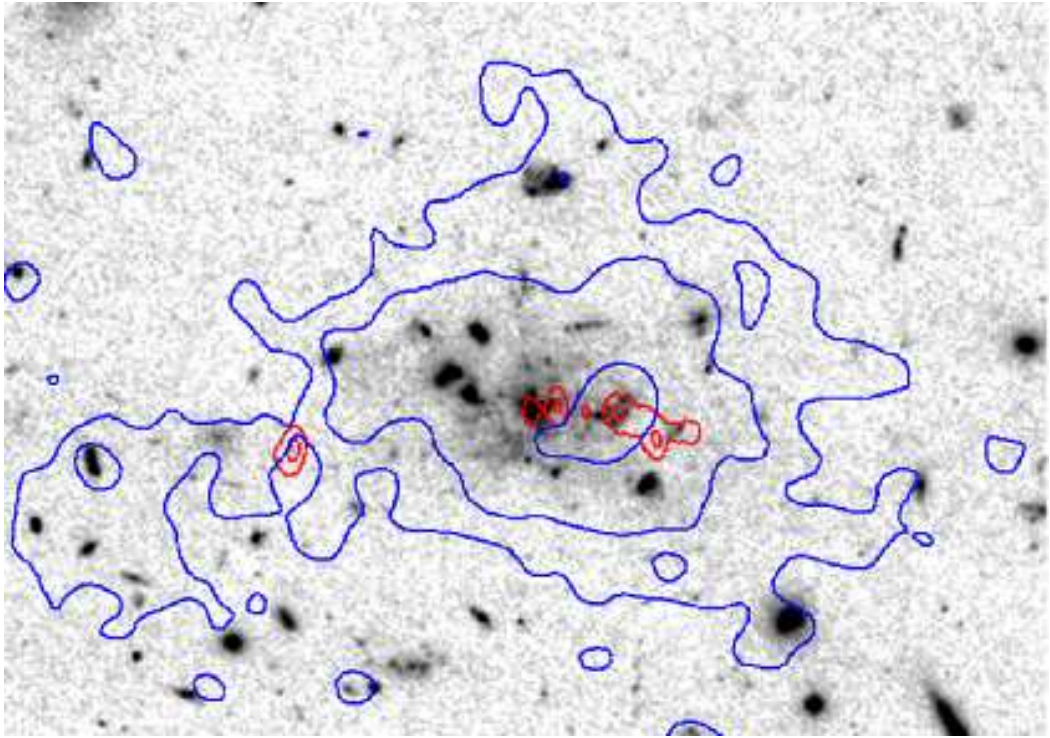}
\end{center}
\caption{VLT Ly$\alpha$\ contours (blue) of the giant emission line nebula surrounding MRC 1138--262, and VLA 8GHz 
radio contours (red) of the non-thermal radio structure superimposed on the composite ($g_{475}$ + $I_{814}$) 
ACS image. The gaseous nebula extends for $>$200 kpc.}
\end{figure}

The galaxy is surrounded by a giant ($\sim200$ kpc) emission line nebula (Fig. 2), similar to the large low surface 
brightness halos of ionized gas seen around other high redshift radio galaxies. These halos could be remnant reservoirs 
of gas from which the massive galaxies are forming, or they could be due to gas swept up by previous episodes of starburst 
superwinds or large-scale AGN outflows (e.g. Haiman et al. 2000, Taniguchi \& Shioya 2000, Haiman \& Rees 2001, Villar-Mart{\'\i}n et al. 2006). 
When a powerful source of ionizing radiation becomes active (e.g., AGN or starburst) the halo will begin to emit luminous line emission, 
and high velocities of gas can be observed in regions where it is perturbed by interaction with the radio structure. The size of the 
halos is comparable to that of the diffuse envelopes of cD galaxies in the local Universe, but it is not clear whether these are 
related phenomena.

\section{Discussion}

We interpret the new HST image as showing hierarchical merging in a proto-BCG. The morphological complexity of the MRC 1138--262 
system agrees, at least qualitatively, with predictions of massive galaxy formation (e.g. Dubinsky 1998, Gao et al. 2004). 
The extensive merging provides a plausible mechanism for fueling the radio source, which may subsequently interact with its 
surrounding medium to regulate its own growth by preventing gas accretion from the larger halo when the source is `on' 
(Croton et al. 2006). Evidence for this kind of coupling between the radio structure and the surrounding medium has been 
revealed through recent integral field spectroscopy of the gaseous halo surrounding MRC 1138--262 (Fig. 2). 
Nesvadba et al. (2006) found very high gas velocities and velocity dispersions associated with the radio structure. 
The energy of the outflow is sufficiently high to expel a significant fraction of the mass in gas during the lifetime of 
the radio source. After the star formation has switched off, the galaxy will continue to grow through merging (De Lucia \& Blaizot 2006). 
However, the total stellar mass contained in the population of satellites is believed to be small compared to the mass of the 
central galaxy ($>10^{11}$ $M_\odot$), estimated from its $K$-band magnitude (Pentericci et al. 1998). This may indicate that 
MRC 1138--262 is an exceptionally rare object that accumulated a large amount of its mass within a relatively short time at $z>2.2$. 
The observations presented here provide a unique testbed for simulations of forming massive galaxies at the centers of clusters 
and protoclusters.


\begin{thebibliography}{}

\bibitem[Bahcall \& Soneira(1983)]{1983ApJ...270...20B} Bahcall, N.~A., \& Soneira, R.~M.\ 1983, \apj, 270, 20 
\bibitem[Blakeslee et al.(2006)]{2006ApJ...644...30B} Blakeslee, J.~P., et al.\ 2006, \apj, 644, 30 
\bibitem[Canalizo \& Stockton(2001)]{2001ApJ...555..719C} Canalizo, G., \& Stockton, A.\ 2001, \apj, 555, 719 
\bibitem[Cowie et al.(1995)]{1995AJ....110.1576C} Cowie, L.~L., Hu, E.~M., \& Songaila, A.\ 1995, \aj, 110, 1576 
\bibitem[Croton et al.(2006)]{2006MNRAS.365...11C} Croton, D.~J., et al.\ 2006, \mnras, 365, 11 
\bibitem[De Breuck et al.(2002)]{2002AJ....123..637D} De Breuck, C., van Breugel, W., Stanford, S.~A., R{\"o}ttgering, H., Miley, G., \& Stern, D.\ 2002, \aj, 123, 637 
\bibitem[De Lucia \& Blaizot(2006)]{db06} De Lucia, G., \& Blaizot, J.,  2006, \mnras, submitted 
\bibitem[Dubinski(1998)]{1998ApJ...502..141D} Dubinski, J.\ 1998, \apj, 502, 141 
\bibitem[Dunlop \& Peacock(1990)]{1990MNRAS.247...19D} Dunlop, J.~S., \& Peacock, J.~A.\ 1990, \mnras, 247, 19 
\bibitem[Ford et al.(1998)]{1998SPIE.3356..234F} Ford, H.~C., et al.\ 1998, \procspie, 3356, 234 
\bibitem[Gao et al.(2004)]{2004ApJ...614...17G} Gao, L., Loeb, A., Peebles, P.~J.~E., White, S.~D.~M., \& Jenkins, A.\ 2004, \apj, 614, 17 
\bibitem[Haiman et al.(2000)]{2000ApJ...537L...5H} Haiman, Z., Spaans, M., \& Quataert, E.\ 2000, \apjl, 537, L5 
\bibitem[Haiman \& Rees(2001)]{2001ApJ...556...87H} Haiman, Z., \& Rees, M.~J.\ 2001, \apj, 556, 87 
\bibitem[Heckman et al.(1986)]{1986ApJ...311..526H} Heckman, T.~M., Smith, E.~P., Baum, S.~A., van Breugel, W.~J.~M., Miley, G.~K., Illingworth, G.~D., Bothun, G.~D., \& Balick, B.\ 1986, \apj, 311, 526 
\bibitem[Kaiser(1984)]{1984ApJ...284L...9K} Kaiser, N.\ 1984, \apjl, 284, L9 
\bibitem[Kauffmann \& Haehnelt(2000)]{2000MNRAS.311..576K} Kauffmann, G., \& Haehnelt, M.\ 2000, \mnras, 311, 576 
\bibitem[Kurk et al.(2000)]{2000A&A...358L...1K} Kurk, J.~D., et al.\ 2000, \aap, 358, L1 
\bibitem[Kurk et al.(2004)]{2004A&A...428..817K} Kurk, J.~D., Pentericci, L., Overzier, R.~A., R{\"o}ttgering, H.~J.~A., \& Miley, G.~K.\ 2004, \aap, 428, 817 
\bibitem[Miley et al.(2004)]{M2004} Miley, G.~K., et al.\ 2004, \nat, 426, 47
\bibitem[Miley et al.(2006)]{M2006} Miley, G.~K., et al.\ 2006,  \apjl, 650, L29
\bibitem[Mullis et al.(2005)]{2005ApJ...623L..85M} Mullis, C.~R., Rosati, P., Lamer, G., B{\"o}hringer, H., Schwope, A., Schuecker, P., \& Fassbender, R.\ 2005, \apjl, 623, L85 
\bibitem[Pentericci et al.(1997)]{1997A&A...326..580P} Pentericci, L., R\"ottgering, H.~J.~A., Miley, G.~K., Carilli, C.~L., \& McCarthy, P.\ 1997, \aap, 326, 580 
\bibitem[Nesvadba et al.(2006)]{n06} Nesvadba, N., et al.\ 2006, \apj, in press
\bibitem[Overzier et al.(2006a)]{2006ApJ...637...58O} Overzier, R.~A., et al.\ 2006a, \apj, submitted (astro-ph/0601223)
\bibitem[Overzier et al.(2006b)]{2006ApJ} Overzier, R.~A., et al.\ 2006b, \apj, 637, 58 
\bibitem[Pentericci et al.(1998)]{1998ApJ...504..139P} Pentericci, L., R\"ottgering, H.~J.~A., Miley, G.~K., Spinrad, H., McCarthy, P.~J., van Breugel, W.~J.~M., \& Macchetto, F.\ 1998, \apj, 504, 139 
\bibitem[Pentericci et al.(2000)]{2000A&A...361L..25P} Pentericci, L., et al.\ 2000, \aap, 361, L25 
\bibitem[Pentericci et al.(2001)]{2001ApJS..135...63P} Pentericci, L., McCarthy, P.~J., R{\"o}ttgering, H.~J.~A., Miley, G.~K., van Breugel, W.~J.~M., \& Fosbury, R.\ 2001, \apjs, 135, 63 
\bibitem[Springel et al.(2005)]{2005Natur.435..629S} Springel, V., et al.\ 2005, \nat, 435, 629 
\bibitem[Stanford et al.(2006)]{2006ApJ...646L..13S} Stanford, S.~A., et al.\ 2006, \apjl, 646, L13 
\bibitem[Straughn et al.(2006)]{2006ApJ...639..724S} Straughn, A.~N., Cohen, S.~H., Ryan, R.~E., Hathi, N.~P., Windhorst, R.~A., \& Jansen, R.~A.\ 2006, \apj, 639, 724 
\bibitem[Taniguchi \& Shioya(2000)]{2000ApJ...532L..13T} Taniguchi, Y., \& Shioya, Y.\ 2000, \apjl, 532, L13 
\bibitem[Taniguchi \& Shioya(2001)]{2001ApJ...547..146T} Taniguchi, Y., \& Shioya, Y.\ 2001, \apj, 547, 146 
\bibitem[Venemans et al.(2002)]{2002ApJ...569L..11V} Venemans, B.~P., et al.\ 2002, \apjl, 569, L11 
\bibitem[Venemans et al.(2004)]{2004A&A...424L..17V} Venemans, B.~P., et al.\ 2004, \aap, 424, L17 
\bibitem[Venemans et al.(2005)]{2005A&A...431..793V} Venemans, B.~P., et al.\ 2005, \aap, 431, 793 
\bibitem[Villar-Mart{\'{\i}}n et al.(2006)]{2006MNRAS.366L...1V} Villar-Mart{\'{\i}}n, M., et al.\ 2006, \mnras, 366, L1 
\bibitem[West(1994)]{1994MNRAS.268...79W} West, M.~J.\ 1994, \mnras, 268, 79 

\end{thebibliography}
\end{document}